# From jet algorithms to *C*-algebra.
## Measurement errors and regularization of cuts[1]

Fyodor V. Tkachov

Institute for Nuclear Research of Russian Academy of Sciences,
60th October Ave. 7a, Moscow 117312 Russia

Error enhancement properties of data processing algorithms in elementary particle physics measurements are discussed. It is argued that a systematic use of continuous weights instead of hard cuts may reduce errors of the results at the cost of a marginal increase of computer resources needed.

At the two previous workshops of this series I discussed a new theoretical scheme for describing multijet structure of hadron events at high energies. Some of the first versions presented then proved to be too crude and were not included into the published talks [1], [2]. However, a rather complete — and, hopefully, reasonably final — account of the theory has been made public in [3], [4]. The theory of [3], [4] is based, among other things, on some simple arguments about error-enhancement properties of data processing algorithms in high energy physics experiments. Those arguments are by no means limited to jet-related problems and deserve to be clarified in a broader perspective. In fact, they seem to be worth to have in view when processing data in any elementary particle physics experiments where cuts with respect to a parameter are routinely employed for event selection and, on the other hand, numerical quality of results is at stake (precision measurements and low-signal situations). On the other hand, a clear understanding of those arguments seems to be essential for a smooth and correct transition to the use of the formalism of *C*-algebra in jet-related problems. This is what I would like to discuss in the present talk.

## *C*-algebra. An overview                                                1

<u>*C-algebra*</u>[2] is a convenient label to refer to the new formalism for measuring multijet structure (what can be roughly described as 'jet counting') [4]. It concerns the processing of data obtained from calorimetric detector installations of high energy physics where, instead of individual hadrons produced in an event, one focuses on narrowly collimated sprays of hadrons (jets). The conventional methods are based on the so-called jet definition algorithms, which work as follows:

$$\mathbf{P} = \{p_1, p_2, \ldots p_{N_{\text{part}}}\} \xrightarrow{\text{jet finding algorithm}} \mathbf{Q} = \{Q_1, Q_2, \ldots Q_{N_{\text{jets}}}\}, \qquad 1.1$$

---

[1] Talk at the X Int. Workshop on High Energy Physics. Zvenigorod, Russia, 19-26 September 1995. Moscow: INP MSU, in print.

[2] *C* from 'calorimetric'. The term has nothing to do with the *C\**-algebras well-known in mathematics, as 'topology of event' has nothing to do with what mathematicians call 'topology'.

where $p_i$ are the 4-momenta of the particles in the final state while $Q_j$ are the 4-momenta of the resulting jets. In other words, the input of a jet algorithm is an event **P** as seen by the calorimetric detector[3] while the output **Q** consists of the jets' 4-momenta (it is convenient to call **Q** <u>*jet pattern*</u> of the event[4]). While the event **P** may have $O(100)$ particles, a majority of events have only a few jets. Therefore, the jet pattern **Q** has typically much fewer 4-momenta than the original event **P** and can be studied much easier.

'To study multijet structure' means to compute various characteristics like the fraction of events with a given number of jets in a given reaction; the distributions of invariant masses of multijet substates; etc. Needless to say the jet pattern **Q** depends on how one defines jets. There is no unique criterion for that (the so-called problem of ambiguities of jet algorithms; cf. e.g. [12]). This results in systematic errors which may be so large as to limit the expected precision of results. For instance, the results of Monte Carlo modeling of the top mass measurements at the LHC [13] indicate that the dominant source of systematic errors will be the ambiguities of jet definition.

The formalism of *C*-algebra is intended as a replacement for the procedures based on jet definition algorithms in situations when the ambiguities of the latter are non-negligible. Let me first briefly explain (i) what the *C*-algebra is; (ii) what sort of restrictions went into its construction.

*What is the C-algebra?* The *C*-algebra proper consists of:

• A family of shape observables of a special form (the so-called *C*-correlators) that form a sort of basis for the *C*-algebra. Among the *C*-correlators, the so-called jet-number discriminators [3] play a special role. Other examples of *C*-correlators are the energy, invariant mass, transverse energy of a multiparticle (sub) state.

• A few rules to construct new observables that, by definition, also belong to the *C*-algebra (which means they possess some nice properties described below). The least trivial rule is the spectral construction that allows one to write down observables that measure mass spectra of multijet substates and other similar quantities.

• A few rules to translate properties of multijet structure from the language of jets into the language of *C*-algebra. For instance, instead of the $n$-jet fractions used in the conventional approach, here we talk about average values of the jet-number discriminators (which also form a sequence $\langle \mathbf{J}_m \rangle$, $m = 1, 2, \ldots, +\infty$ similar to $\sigma^{n \text{ jets}}$, $n = 1, 2, \ldots, +\infty$). The value of the $m$-th jet number discriminator $\mathbf{J}_m(\mathbf{P})$ is naturally interpreted as a "weight" of the $\geq m$-jet component in the event **P**. One can also write down formulas that have the same physical meaning (and roughly the same numerical content[5]) as, say, the invariant mass distribution of 3-jet substates. As the examples given in [4] demonstrate, the expressive power of *C*-algebra is sufficient for expressing physics in situations of practical interest.

• Computational methods for extracting the values of the *C*-observables from data. They exploit the properties of the observables from *C*-algebra. The most important one is probably the so-called optimal preclustering [15], [4] which is similar to the conventional recombination algorithms but whose criterion is fixed uniquely by the requirement of minimizing the approximation errors due to replacement of particles of the exact final state by pseudoparticles (protojets). But other tricks (expansion in energies of soft particles [4]) also deserve attention.

---

[3] Non-calorimetric information can be taken into account as 'external parameters'; see [4] for more details.

[4] It is rather characteristic of the conceptually amorphous conventional jet-counting paradigm that such a central object as jet pattern does not even seem to have been given a distinctive name.

[5] "Roughly" because if the new formalism is to be in any sense better than the conventional approach it cannot yield exactly the same numbers.

I emphasize that the observables from the *C*-algebra are defined directly in terms of the unprocessed event (the l.h.s. of 1.1) rather than in terms of the jet pattern (the r.h.s. of 1.1). Yet by construction, the qualitative content of the *C*-observables is the same as that of the conventional observables defined via jets. Jet algorithms retain the role of an approximation trick with a well-defined criterion, namely, the numerical quality of the resulting approximation. The approximation error is proportional to $y_{\text{cut}}$ (the jet resolution parameter of the conventional algorithms). Then what is known as ambiguities of jet algorithms becomes an unavoidable ambiguity of an approximation and is eliminated by letting $y_{\text{cut}} \to 0$. To do this one needs more computing power; but by the time, say, the LHC begins operation the computing power will become many times cheaper so the problem mentioned in [13] can be expected to be alleviated.

*The considerations that went into the construction of C-algebra* are as follows [4]:

- Optimal stability of the observables with respect to data errors [1]. "Optimal" means that for a given physical feature one attempts to construct an observable that is least sensitive to data errors and statistical fluctuations.

- Compatibility with quantum field theory. The latter simply means that the observables one uses should conform to the general structure of QFT because the latter is the most fundamental mechanics of elementary particles. At a practical level such a conformance would mean more precise and systematic theoretical predictions.

The QFT aspect of the *C*-algebra is discussed by N.A. Sveshnikov at this workshop [14]. I would like to concentrate on the first aspect.

The mapping described by 1.1 can be regarded as a function $f(\mathbf{P})$ on final states $\mathbf{P}$ whose mathematical structure (the mathematical nature of its values) is rather complex. As a first step, consider the following question: how large are variations of the values of such a function given variations of its argument? Indeed, the event $\mathbf{P}$ is never seen precisely, there are always measurement errors. Which functions $f(\mathbf{P})$ are least sensitive to such errors? Ref. [4] contains a systematic answer to this question. Briefly, the answer is that the functions $f(\mathbf{P})$ should be continuous in a special sense. Indeed, since at high energies the number of particles on the l.h.s. of 1.1 is not limited, the mathematical space of the events $\mathbf{P}$ should be regarded as infinitely dimensional, and in infinitely dimensional spaces many radically non-equivalent continuities are possible. The one that corresponds to minimizing the effect of detector errors in calorimetric experiments is determined uniquely by the 'kinematical' structure of such detectors and was called *C-continuity* [1]–[4]. It turns out that the *C*-correlators that form the basis of the *C*-algebra are exactly the simplest *C*-continuous functions while the rules of *C*-algebra for constructing new observables are chosen such as to preserve the property of *C*-continuity.

It should be emphasized that the restriction of *C*-continuity is a very general one and concerns *only* sensitivity to data errors, and the rather tight "axiomatic" construction of *C*-algebra in [4] with all assumptions explicitly stated, is intended to ensure that no physics is lost when one uses the *C*-algebra to describe the multijet structure. (What one does loose by restricting to *C*-continuous observables is errors and the naive visual simplicity of the jet pattern.)

To put it shortly, the formalism of *C*-algebra is a precise mathematical expression at a "syntactic" level of an essentially physical requirement of optimal stability of measurements with respect to errors always present in the data from real-world finite-precision detectors.

## One-dimensional toy model 2

Let us concentrate on the requirement of stability of measurements with respect to errors in a most general context. Namely, it will be seen that it is advantageous to always use continuos observables (regularized cuts) instead of hard cuts habitually employed by experimentalists for event selection. "Always" means practically in all high-precision/low signal situations unless the additional cost of data processing is clearly prohibitive.

In addition to the stability with respect to data errors discussed in [4] I would like to point out that the continuity of observables reduces also the purely statistical fluctuations due to limitation of a finite event sample. This will be discussed in Sec. 4.

Suppose one deals with a collection of events **P** corresponding to a quantum reaction, and one classifies those events according to one number which therefore represents the event. Then the continuum of events can be represented by a segment of the real axis which for simplicity can be taken to be [0,1].

The events are produced with a (unknown) probability measure $\pi(\mathbf{P})$ determined by the amplitudes of the process (*S*-matrix elements squared). Each observable *f* is defined by a function on the events, $f(\mathbf{P})$. The average value of the observable, $\langle f \rangle$ is the integral of $f(\mathbf{P})$ against the measure $\pi(\mathbf{P})$:

$$\langle f \rangle \equiv \int_0^1 d\mathbf{P}\, \pi(\mathbf{P})\, f(\mathbf{P}) \,. \qquad 2.1$$

In practice one deals with a finite selection of events $\mathbf{P}_i$, $i = 1,\ldots,Z$. So Eq. 2.1 is replaced by

$$\langle f \rangle \approx Z^{-1} \sum_{i=1}^{i=Z} f(\mathbf{P}_i) \,. \qquad 2.2$$

Such a replacement results in an error which we will refer to as *statistical error* and consider in Sec. 4. The other kind of error is due to the fact that the position of each event **P** on the real axis is known with an error (such errors are referred to here as *data errors*). This means that instead of events $\mathbf{P}_i$ we deal with distorted events $\mathbf{P}_i^\varepsilon$, where the parameter $\varepsilon$ describes the size of the data errors. Therefore, the expression one deals with in the final respect is as follows:

$$\langle f \rangle \approx Z^{-1} \sum_{i=1}^{i=Z} f(\mathbf{P}_i^\varepsilon) \,. \qquad 2.3$$

We wish to study how the properties of *f* affect the difference between the exact value 2.1 and the measured value 2.3.

Before we proceed it is useful to note that there are two different cases one should have in view. In practice they never occur in a pure form but it is useful to understand them separately:

(i) The first case is when one selects events that fall, say, above the cut $\mathbf{P}_{\text{cut}}$, and counts the number of events thus selected. Presumably the events thus selected carry some physical fea-ture one wishes to study (e.g., events corresponding to a production of a new particle may predominantly occur at **P** above the cut). The resulting observable is expressed by the integral

$$\int_{\mathbf{P}_{cut}}^{1} d\mathbf{P}\, \pi(\mathbf{P}) = \int_{0}^{1} d\mathbf{P}\, \pi(\mathbf{P})\, \Theta_{cut}(\mathbf{P}) \equiv \langle \Theta_{cut} \rangle, \qquad 2.4$$

where

$$\Theta_{cut}(\mathbf{P}) = 1 \text{ for } \mathbf{P} > \mathbf{P}_{cut} \text{ and } 0 \text{ otherwise.} \qquad 2.5$$

In this case one deals with a hard cut represented by a discontinuous step function.

(ii) The other case is when the observable is a smooth function. "Smooth" implies that

$$\left| f(\mathbf{P}_i^\varepsilon) - f(\mathbf{P}_i) \right| = O(\varepsilon). \qquad 2.6$$

Note that the latter property is violated for 2.5 because infinitesimal variations of the argument **P** around the cut $\mathbf{P}_{cut}$ result in oscillations of the value $f(\mathbf{P})$ between 0 and 1.

## Stability with respect to data errors 3

Let us consider the effect of data errors on the observable $\langle f \rangle$. The difference between the cases (i) and (ii) can be shown as follows:

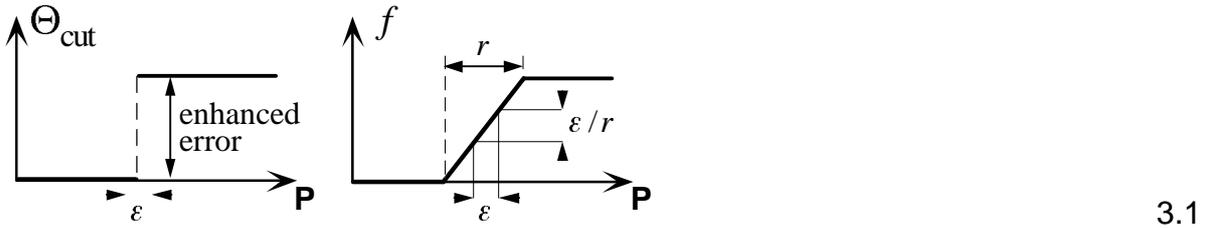

3.1

We have chosen $f$ to be piece-wise linear, with the non-constant part placed around the position of the step in the left figure. Thus the weight in the right figure can be regarded as a *regularization* of the hard cut in the left one. This can also be referred to as *regularized cut*. The interval of the linear interpolation in the right figure is the *regularization interval*.

The qualitative difference of the two pictures 3.1 translates into a difference between how the measurement errors from many events accumulate when the integration is performed. The key point here is that measurement errors for different events are independent, so it is the corresponding $\sigma^2$ that are added rather than $\sigma$ (should the latter be true, the discontinuity would have played no role). As a result, the integral error is quite different in the two cases.

For simplicity suppose $\pi(\mathbf{P}) \sim \text{const}$, so that the generated events are distributed more or less uniformly. Then the measurement errors for events that occur close to the cut (their fraction is $O(\varepsilon)$) will induce fluctuations of order $O(1)$ for the value of $\Theta_{cut}(\mathbf{P})$ because of the discontinuity. Then the variance of the error induced in the result is

$$\sigma_{hard}^2 = O(\varepsilon) \times [O(1)]^2 = O(\varepsilon). \qquad 3.2$$

The integral error for the case (ii) is due to events **P** for which $f(\mathbf{P})$ is computed with errors due to errors in their position on the horizontal axis. Obviously, these are the events that fall into the regularization interval. The fraction of such events is $O(r)$. The variance of the value of $\Theta_{\text{cut}}$ for each such event is

$$\left[ O(\varepsilon) \times \tfrac{1}{r} \right]^2 = O\!\left(\tfrac{\varepsilon^2}{r^2}\right). \tag{3.3}$$

Since measurement errors for different events are independent, the total variance is estimated by Eq. 3.3 times the fraction of such events, $O(r)$, i.e.

$$\sigma^2_{\text{reg}} = O\!\left(\tfrac{\varepsilon^2}{r}\right). \tag{3.4}$$

For $r = O(\varepsilon)$, this degenerates into 3.2, as expected.

The net effect is that the resulting error interval in the regularized case is suppressed as compared with the hard cut case by

$$\frac{\sigma_{\text{hard}}}{\sigma_{\text{reg}}} = O\!\left(\sqrt{\tfrac{r}{\varepsilon}}\right) \tag{3.5}$$

A conclusion is that the slower the variation of the observable over the entire continuum of the events, the smaller is the effect of the data errors.

## Stability with respect to statistical fluctuations 4

This issue has not been addressed in [4] but it is quite remarkable that the statistical fluctuations are also suppressed in the case of regularized cuts. (Here we neglect the data errors for simplicity.) This is seen as follows. The variance of the result is, roughly,

$$\sigma^2_{\langle f \rangle} \approx \frac{1}{N} \int_0^1 d\mathbf{P}\, \pi(\mathbf{P}) [f(\mathbf{P}) - \langle f \rangle]^2. \tag{4.1}$$

Plot the integrand for the cases shown in 3.1:

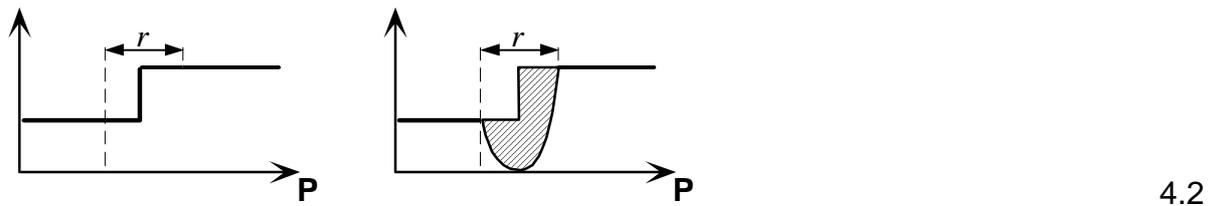

(4.2)

One can see from the right figure that the integrand in the regularized case takes a parabolic dip around the step of the non-regularized case. So the regularization cuts out a chunk (the hatched region in the right figure) from the variance as compared with the hard cut. Note that the resulting suppression factor is independent of $N$ for large $N$; cf. how $N$ enters 4.1.

This effect may sometimes turn out more important numerically than the suppression of data errors discussed in the preceding Sec. 3.

## "Physical meaning" of regularized cuts 5

From the point of view of errors it is clear that regularizing the cuts results in a suppression of both statistical errors due to finite event samples and the errors induced by errors in the unprocessed data. Many people would, however, ask about physical meaning of the regulariza-tion of cuts. The answer to such a question[6] is that however much or little physical meaning regularized cuts may have, hard cuts have even less of it in the first place. Indeed, there is no intrinsic boundary in the continuum of events that separates, say, the events due to top decay from QCD background. At a crude (qualitative) level both hard and regularized cuts work roughly the same. When one starts worrying about errors of the results, the physical focus shifts to the issue of quality of measurements, and a regularization becomes obligatory.

One can also say that various observables one uses represent the vision mechanism with which to observe nature; then the regularization of cuts is a fine tuning of that mechanism which is necessary in high-precision/low-signal situations.

One may also interpret a regularized cut as a weight of the corresponding physical feature in a given event[7]. Then summation over all events yields the total weight of the feature in the entire event sample. Or the weight can be thought of as a measure of difference of the event from (or its similarity to) some "exemplar events" that are chosen as perfect representatives of the feature one wishes to quantify. Then summing over all events yields the average "distance" of the event sample from the exemplar events.

## Uses of regularizations 6

The jet-number discriminators and the *C*-algebra defined and studied in [4] are rather sophisticated examples of how one can construct regularized weights describing various jets-related physical features. I would like to emphasize here that in many situations one simply needs to perform a cut with respect to one parameter. Then it would be sufficient to regularize the cut using a weight of the form shown in the right figure 3.1. Let us briefly discuss this.

First recall that the exact position of hard cut is chosen to maximize the signal/background ratio. One sees that regularization introduces an additional parameter — the width of the regularization interval which is also determined from the same condition. It is perfectly clear that an additional parameter cannot make the situation worse because the case of hard cut is restored for $r \to 0$. Clearly the situations when the optimum is reached for $r = 0$ are exceptional. In general the signal/noise ratio is maximized for some $r > 0$.

Of course, both theoretical calculations and the processing of data should be performed with the same cut. Note also that matching theoretical predictions with data for high-precision extraction of parameters would require a closer interaction of theorists and experimentalists.

A generalization to the case of more than one parameter with respect to which to perform a cut is straightforward. Consider e.g. the procedure described in [16] where a techniques of neural networks was used to determine an optimal position of a curvilinear cut in the 2-dimensional parameter space. It is interesting to note that, on the one hand, a regularization can be applied to curvilinear cuts as well as to the cuts with respect to each of the two parameters. On the other hand, the use of regularization should reduce (and sometimes eliminate) the need in complex cuts especially when the regularization region is wide enough because then the results are less sensitive to the details of the cut's geometry.

For the sake of illustration I would like to point out that regularization can be applied to cuts used in the conventional jet definition algorithms. For instance, a standard definition of the jet cone corresponds to using a step function that is equal to 1 within the cone and 0 outside. One can regularize it by replacing it with a continuous weight (that linearly interpolates between 0 and 1 with respect to the angular distance from the jet center, in an interval around what would normally be the jet's radius). Then if a particle happens to be in the regularization region, only its fraction (proportional to the corresponding weight value) is to be included into the jet. This trick alone would somewhat stabilize the "ambiguities" (= instabilities) of jet algorithms and reduce sensitivity of their results to details of jet definition.

---

[6] Apart from the abuse of the term "physical meaning", which is not discussed here.

[7] Such an interpretation was influenced by a probabilistic interpretation of jet-number discriminators due to F. Dydak. (Probabilities here are, of course, only a figure of speech; there is nothing stochastic about such weights.)

A still better way, of course, is to systematically use the formalism of *C*-algebra.

Another point worth bearing in mind is that a cut is a ubiquitous concept in experimental data acquisition systems. Clearly, a regularization of those cuts may be also useful. The simplest way to achieve this is at the off-line stage of data processing. Then each event that is close to the cut should be taken with a weight.[8]

## Splines instead of bins 7

A more complex example of a regularization of cuts concerns computation of various distributions of events with respect to one or more parameters. Indeed, a conventional approach is to split the parameter's interval of variation into several subintervals (bins), to compute the value of the parameter for each event, and to put the event into the corresponding bin. One thus obtains a histogram (or a Lego plot if two parameters are involved).

Such a procedure is equivalent to using a sequence of hard cuts. In view of the above discussion of cuts' regularization, an obvious way to improve upon it is to regularize each of the cuts involved. This can be achieved using linear splines instead of bins as described in [4].

Concerning the version given in [4], a remark is in order. In [4], the distributions are generated not for events as such but rather for quantities that are computed from the events. In particular, each event is processed in such a way (involving integrations over spheres) that it results in many "units of weight" scattered over the interval of variation of the parameter. The simpler scheme outlined below deals directly with the distribution of events, without assuming an intermediate step as in [4]. However, some details of mathematical interpretation discussed in [4] (esp. sec. 15) are not repeated here.

Let $s$ denote the parameter with respect to which one wishes to compute the distribution of events. Suppose for simplicity it varies in the unit interval $[0,1]$. Denote the exact distribution as $\rho(s)$. We assume here that $\rho(s)$ is a continuous function. This is always the case when one computes a distribution directly from events. Consider the following continuous function:

$$\rho_N(s) \equiv \int_0^1 ds' \, \rho(s') h_N(s' - s), \qquad 7.1$$

where the sequence $h_N$ converges to the $\delta$-function (concrete examples are given below):

$$\lim_{N \to \infty} h_N(s) = \delta(s). \qquad 7.2$$

---

[8] A fancier way that requires a modification of the data acquisition system might be to compute the weight on-line prior to the selection. Then the event is selected for writing to tapes with a probability depending on (e.g. equal to) the computed weight (because contributions of the "marginal" events would be suppressed by the weight anyway). This might allow one to probe a somewhat broader region in the continuum of all events without increasing the number of events stored to tapes.

It is clear that

$$\lim_{N \to \infty} \rho_N(s) = \rho(s) \qquad 7.3$$

for each $s$.

The point is, it is only quantities of the form 7.1 that one can compute, never $\rho(s)$ itself (cf. the discussion of Sec. 2). $\rho(s)$ can only be restored after taking the limit $N \to \infty$. The problem therefore is how to choose the weight $h_N$ so as to optimize the convergence.

First of all note that given a finite sample of events, $\rho_N(s)$ for each $s$ can be computed from the following formula

$$\rho_N(s) \approx r_N(s) \equiv Z^{-1} \sum_e h_N(s_e - s), \quad Z = \sum_e 1, \qquad 7.4$$

where the index $e$ enumerates events, and the normalization factor $Z$ is the number of all events of the sample.

A standard bin-type scheme is reproduced as follows. One chooses $h_N(s) = 1$ or $0$ depending on whether or not $|s| < \Delta s/2$ (where $\Delta s = 1/N$) and computes $r_N(s_i)$ for a finite number of equidistant values $s_i$, $i = 0, \ldots N$; $s_0 = 0$ and $s_N = 1$. In practice one runs the loop over $i$ and the loop over $e$ in a different order: for each event $e$ one determines $i$ for which the event gives a non-zero contribution to $r_N(s_i)$ (i.e. which "bin" the event falls into) and modifies $r_N(s_i) \leftarrow r_N(s_i) + 1$. In the end one normalizes $r_N(s_i)$ by the total number of events.

The above reinterpretation of the standard scheme makes it obvious how it should be modified to incorporate regularization. It is sufficient to replace the above discontinuous $h_N$ with a "regularized" one (which should also satisfy 7.2). The simplest choice is as follows:

$$h_N(s) = N h(Ns), \quad h(s) = \max(0, 1 - |s|). \qquad 7.5$$

With this definition, one computes — as in the bin-type scheme — the array of values $r_N(s_i)$ from the formula 7.4. Again, for the sake of efficiency one runs the loops in the opposite order: One starts with $r_N(s_i) \equiv 0$ for all $i$, and runs a loop over all events. For each event one determines the corresponding $s_e$, and then one determines $i$ such that $s_i \leq s_e \leq s_{i+1}$. Then one redefines

$$r_N(s_i) \leftarrow r_N(s_i) + \frac{s_{i+1} - s_e}{\Delta s},$$

$$r_N(s_{i+1}) \leftarrow r_N(s_{i+1}) + \frac{s_e - s_i}{\Delta s}. \qquad 7.6$$

After the loop is completed, one normalizes the array $r_N(s_i)$ by the total number of events.

Let us now turn to the important issue of the optimal choice of the regularization parameter $N$ (or, equivalently, $\Delta s$). Note that this problem is analogous to that of choosing the width of the regularizing interval $r$ in Secs. 2–4.

First of all, one can simply apply the same criteria to choose $N$ as used in the case of the bin-type scheme (irrespective of whether or not they are scientifically justified). However, there are more specific considerations to take into account.

Because of 7.3, one would like to choose $N$ as large as possible. This, however, would in-terfere with the error suppression mechanisms discussed in Secs. 2–4. But too small $N$ would result in a loss of information. Therefore, one should choose an optimal $N$ in order to achieve a balance. From the analysis of Sec. 3 one can see that $\Delta s$ should be larger by a factor $\geq 2$ than the uncertainty in $s_e$ due to data errors. A suppression of statistical fluctuations discussed in Sec. 4 is harder to estimate. Therefore in practice one should perform a preliminary MC modeling or employ some heuristic procedure.

One may find useful the following "folding" trick that allows one to halve $N$ (equivalently, to double $\Delta s$) without straightforward recalculation of $r_N(s_i)$. The trick is based on the identity

$$h_N(s - s_i) = \tfrac{1}{4} h_{2N}(s - s_{2i-1}) + \tfrac{1}{2} h_{2N}(s - s_{2i}) + \tfrac{1}{4} h_{2N}(s - s_{2i+1}) \,, \qquad 7.7$$

which translates into the following folding formula:

$$r_N(s_i) = \tfrac{1}{4} r_{2N}(s_{2i-1}) + \tfrac{1}{2} r_{2N}(s_{2i}) + \tfrac{1}{4} r_{2N}(s_{2i+1}) \,. \qquad 7.8$$

Then one should compute $r_N(s_i)$ once for sufficiently large $N$ (e.g. corresponding to $\Delta s \lesssim$ the uncertainty in $s_e$ due to data errors); the chosen $N$ should allow a sufficient number of divisions by 2 to allow as many foldings as desired. Performing foldings one may be able to determine an optimal value for $N$.

One could use higher order splines (corresponding to using smoother weights than the piecewise-linear ones considered above) in a similar manner to achieve an even better control of errors. The necessary modifications of formulas are straightforward.

## Discussion 8

Psychologically, it is important to explain why regularized weights are rarely considered in standard treatments of statistical methods. First, neither the effects of data errors in the above sense nor precision measurements play a noticeable role in standard applications from which mathematical statistics evolved (such as demography). Second, there is a certain inertia of thought due to the fact that the conventional Kolmogorov axiomatics of the theory of probability is based on the old-fashioned theory of measure with measures defined as additive functions on subsets (the measure of a subset is the same as the integral of the function that takes the value 1 on the subset and 0 outside it).

On the other hand, the construction of $C$-algebra is based on the modern functional-analytic approach (cf. [17]) in which a measure is first defined as an integral of continuous functions (e.g. via the Riemann sums) and then extended to other functions (e.g. discontinuous functions such as characteristic functions of subsets) via continuity. The two variants of the theory of measure are largely equivalent: the definitions of one become theorems of the other and vice versa. However, the two approaches differ appreciably in the heuristic aspect and lead one to consider different problems. Note that the functional-analytic approach is more immediately relevant to the subject of calorimetric measurements.[9]

Note also that continuous regularizations are routinely used in other branches of applied mathematics (cf. the body of theoretical and applied research conducted by Tikhonov and his school [18]). The problems dealt with above can actually be regarded as rather simple special cases of the more general mathematical problems considered by Tikhonov et al.

---

[9] Note that the first variant of the theory described in [1] used, by inertia, the old-fashioned approach with elementary detector modules associated with subsets of the unit sphere, which caused difficulties until the matter was clarified in [3] where the correct treatment in terms of continuous functions was found.

It is clear that the regularization of cuts is a useful option in high-precision statistical measurements and deserves to be studied and used in a systematic manner whenever appropriate (which seems to include practically all precision measurements in high energy physics).

*Acknowledgements*   This work was made possible in parts by the International Science Foundation (grants MP9000/9300) and the Russian Fund for Basic Research (grant 95-02-05794).

## *Notes added*

**1.**

In [4], Sec. 10, the definition of "elementary mass detector", eq. 10.3, should, perhaps, be slightly modified by making the regularization interval independent of $R$ (indeed, it only ought to depend on the detector module sizes). This modification does not affect any of the general properties of the spectral discriminators with respect to sensitivity to masses of $n$-jet substates.

Recall that the size $r$ of the regularization interval should be $\gtrsim 2\varepsilon$ where $\varepsilon$ is the size of the measurement errors (cf. 3.1–3.5 above and sec. 2.8 in [4]). In the case of 10.3, $\varepsilon$ corresponds to the detector grid step. However, the size of the regularization interval in eq. 10.3 as it stands is $O(R)$. This is not good if $R \lesssim \varepsilon$ (because then the regularization interval is too small for regularization to work), and it is not good for $R \gg \varepsilon$ because then the regularization interval may be unnecessarily large.

To rectify this, one should use a definition like the following one:

$$\Phi_{\hat{q};R}(\hat{p}) = \Phi(\Delta_{qp};R;\varepsilon), \text{ where} \qquad \text{10.3a}$$

$$\begin{aligned}
\Phi(d;R;r) &= 1 && \text{if } d \leq R, \\
&= 0 && \text{if } d \geq R+r, \\
&= \text{linear interpolation in between}, && \text{10.4a}
\end{aligned}$$

where $r$ should be at least $2\varepsilon$ (cf. 3.1–3.5 above) or more if one wishes so.

**2.**

If one computes a histogram replacing bins with linear splines as suggested above in sec. 7 then the overall effect of reduction of the purely statistical component of errors (sec. 4 above) in the resulting distribution can be seen to be at best ~ 3% (which may be barely enough to be worth the trouble: I take as a guide a list of error reductions presented by an experimentalist at a workshop; the list started with a 3% error reduction due to a smart design of a cathode; the cost of implementing a regularization is, of course, negligible in comparison [apart from the mental effort ☺]; also, one usually employs several types of discontinuous cuts so the effect may add up).

Note that the 3% are measured using the $L_2$ norm (least squares method). If one measures the overall error using the $C_0$ norm (max value of pointwise deviations) *then the reduction is more appreciable* (I have no explicit expression for this case to make a numerical estimate). The latter way of measuring the overall error corresponds to the so-called *robust statistical methods*.

However, the reduction of that component of resulting errors which is due to "measurement errors" is governed by the consideration of sec. 3 and may still be significant — depending on a concrete situation.

**3.**

The physical meaning of $R$ in the case of spectral discriminators is not the same as in the case of jet counting. In the latter case, it is, roughly, the jet cone size. In the former case, the jet cone size is the *lower* bound for the interval of $R$ for which the discriminator should be computed (cf. fig. 10.26[right] in [4] where the jet cone size is, roughly, $R_j^-$ ).

**4.**

I emphasize that the optimal jet algorithm described in sec. 7 of [4] *remains the same both in the case of $e^+e^- \to$ hadrons and in the case of hadron-hadron collisions* — all one has to do in the latter case is to replace the energies with transverse energies as described in sec. 13.18 of [4]. Thus the optimal jet algorithm is not only optimal but also *universal*.